\documentclass[english,aps,showpacs,superscriptaddress,nofootinbib,dvipsnames]{revtex4-2}

\usepackage[T1]{fontenc}
\usepackage{float}
\usepackage{amsmath}
\usepackage{txfonts}
\usepackage[normalem]{ulem}
\usepackage{amsfonts}
\usepackage{multirow}
\usepackage{mathrsfs}
\usepackage{subfig}
\usepackage{amsmath}
\usepackage{amssymb}
\usepackage{bm}
\usepackage{bbm}
\usepackage{hyperref}
\usepackage{xcolor}
\usepackage{babel}
\usepackage{graphicx}
\usepackage{caption}
\usepackage{soul}
\setulcolor{red}
\setstcolor{red}

\newcommand {\bseq}{\begin{subequations}}
\newcommand {\eseq}{\end{subequations}}

\newcommand*{\rom}[1]{\uppercase\expandafter{\romannumeral #1\relax}}

\newcommand{\nn}{\nonumber}

\newcommand{\dd}{{\mathrm{d}}}

\newcommand{\abs}[1]{\left\lvert{#1}\right\rvert}
\newcommand{\vb}[1]{\mathbf{#1}}
\newcommand{\pqty}[1]{\left({#1}\right)}
\newcommand{\bqty}[1]{\left[{#1}\right]}

\begin{document}

\title{Deciphering the coalescence behavior of Coulomb-Schr\"odinger atomic wave functions
	from an operator product expansion}

\author{Yingsheng Huang}
\affiliation{Institute of High Energy Physics, Chinese Academy of
	Sciences, Beijing 100049, China}
\affiliation{School of Physics, University of Chinese Academy of Sciences,
	Beijing 100049, China}
\affiliation{Department of Physics and Astronomy, University of Utah, Salt Lake City, UT 84112, USA}

\author{Yu Jia}
\affiliation{Institute of High Energy Physics, Chinese Academy of
	Sciences, Beijing 100049, China}
\affiliation{School of Physics, University of Chinese Academy of Sciences,
	Beijing 100049, China}

\author{Rui Yu}
\affiliation{Institute of High Energy Physics, Chinese Academy of Sciences, Beijing 100049, China}
\affiliation{School of Physics, University of Chinese Academy of Sciences,
	Beijing 100049, China}
\affiliation{School of Physical Science and Technology,
	Inner Mongolia University, Hohhot 010021, China\vspace{0.2cm}}
\affiliation{Beijing computational Science Research Center, Beijing 100049, China\vspace{0.2cm}}

\date{\today}

\begin{abstract}
	We revisit the coalescence behavior of the atomic Schr\"odinger wave functions from the angle
	of an operator product expansion (OPE) within the nonrelativistic Coulomb-Schr\"odinger effective field theory.
	We take the electron-nucleus coalescence as an explicit example to demonstrate our formalism,
	by proving an exact OPE relation to all orders in perturbation theory.
	As a consequence, the celebrated Kato's cusp condition can be readily reproduced.
	Our approach can be readily extended to ascertain the multi-particle coalescence behaviors of atomic wave functions,
	which are otherwise difficult to achieve from the conventional Schr\"odinger equation approach.
	Moreover, by incorporating relativistic effects, our OPE approach appears to be able to provide key insight
	for solving a 90-year-old puzzle concerning the peculiar coalescence behavior of the
	wave function for a Dirac hydrogen atom.
\end{abstract}

\maketitle

\section{Introduction}
Quantum Electrodynamics (QED) is rightfully the underlying theory that reigns all phenomena
in atomic physics and quantum chemistry.
Nevertheless, for the sake of unraveling the property of atoms composed of a heavy nucleus plus $N$ electrons,
it looks practically intractable to directly invoke the relativistic field-theoretical machinery,
rather it turns to be extremely efficient to start from the nonrelativistic
Hamiltonian entailing Coulombic interactions:
\begin{align}
	H_{\rm Coul}= -\sum_{i=1}^N {{\bm\nabla}_i^2\over 2m}-\sum_{i=1}^N {Z \alpha\over r_i}+
	\sum_{j>i=1}^N\frac{\alpha}{r_{ij}},
	\label{Coul:Hamiltonian}
\end{align}
where the electron mass is labeled by $m$, the nucleus and electron carry electric charges of $Ze$ and $-e$,
respectively. $r_i$ signifies the distance between the $i$-th electron and the nucleus, $r_{ij}$ denotes
the distance between the $i$-th and $j$-th electrons. Designing accurate numerical algorithms to solve
the resulting Schr\"odinger equation for a multi-electron atom, $H_{\rm Coul}\Psi=E\Psi$,
was actively pursued shortly after the birth of Quantum Mechanics nearly a century ago,
yet still occupies the central stage of the atomic physics until today~\cite{doi:10.1021/cr200168z}.

Atoms are QED bound states formed by a static heavy nucleus together with some slowly-moving electrons.
In the modern tenet, these electrically-neutral bound states are believed to be best tackled  by
the effective field theory (EFT) dubbed nonrelativistic QED (NRQED)~\cite{Caswell:1985ui}, which directly descends from QED
by integrating out relativistic quantum fluctuations. The notable merit of
the EFT approach is to expedite the systematic inclusion of relativistic corrections for
various atomic and molecular properties~\cite{Pineda:1997ie,Yelkhovsky:2001tx,Brambilla:2017uyf}.

The motif of this work is to demonstrate a remarkable merit of the field-theoretical approach
over the traditional Schr\"odinger equation in portraying the short-distance properties of
atomic wave function. Concretely speaking, in this work we are interested in understanding
the coalescence behavior of atomic wave functions, that is, what kind of universal behavior
the wave function would exhibit when some electrons spatially approach each other, or approach the nucleus.
The knowledge about the correct coalescence behavior of wave functions is important, since they
provide important constraints on the profiles of trial functions. Better trial functions help to
enhance the accuracy of the predicted atomic energy spectrum and various reaction rates. {In addition, recent advancement on direct measurement of wave functions~\cite{Lundeen2011} also grants the opportunity to put this asymptotic behavior on trial. }
We aim to offer a fresh look at this problem from a field-theoretical angle.

First we give a brief account of the history. To our knowledge, the electron-nucleus coalescence behavior of
the hydrogen-like atom with general orbital angular momentum quantum number was first
addressed by L\"owdin in 1954~\cite{Loewdin1954}.
Subsequently, the $S$-wave two-particle coalescence behavior in any multi-electron
atom was summarized by Kato in 1957~\cite{Kato1957},
\begin{align}
	\frac{\partial \Psi}{ \partial r_{12}}\Bigg|_{r_{12}=0}= \gamma\,\Psi(r_{12}=0),
	\label{Kato:cusp:condition}
\end{align}
with $\gamma= {m \alpha\over 2}$ for two-electron coalescence, and $\gamma=-m Z\alpha$ for an electron coinciding
with the nucleus. \eqref{Kato:cusp:condition} is often called Kato's cusp condition.
In the '60s, the cusp condition was extended to the hydrogen molecule~\cite{Kolos:1960zz,Pack1966}.
Recently, the three-particle coalescence behavior for atomic and molecular wave functions has been addressed by  Fournais {\it et al.}~\cite{Fournais2005}.
To date nothing is known about four or more particle coalescence behavior of atomic wave functions.

Most preceding work in this area heavily relies upon complicated mathematics in seeking approximate
solutions of eigen-wave functions of \eqref{Coul:Hamiltonian}, which is not so illuminating
In this work, we attempt to provide an alternative perspective, {\it i.e.},
these {\it universal} coalescence behaviors can actually be best understood within the nonrelativistic
EFT, by invoking the powerful {\it operator product expansion} (OPE) technique.

OPE was originally formulated by Wilson in 1969~\cite{Wilson:1969zs}.
Shortly after it was applied to account for the scaling violation observed in
deep-inelastic scattering experiment, and played a vital role in establishing QCD as the
fundamental theory of strong interaction. OPE has also served a crucial element in the
influential QCD sum rules to extract nonperturbative hadron properties~\cite{Shifman:1978bx}.

Besides its ubiquitous applications in high energy physics, OPE also proves to be useful in
the realm of atomic physics.
It was Lepage who first clarified the implication of renormalization and OPE in
nonrelativistic quantum mechanics~\cite{Lepage:1997cs}.
In 2008, Braaten and Platter applied the OPE technique to prove the Tan relation for the unitary Fermi gas with a zero-range interaction~\cite{Braaten:2008uh}. Recently Hofmann {\it et al.} used OPE to understand the coalescence
behavior of electrons within the jellium model~\cite{Hofmann:2013oia}.

In this work, we apply the OPE to deduce the coalescence behaviors of atomic wave functions.
Unlike \cite{Braaten:2008uh,Hofmann:2013oia}, the nonrelativistic effective theory we start with
has a direct connection to QED.
We will also present a rigorous proof based on perturbation theory.
Our work can be further improved by including more fields in the operator product,
as well as including relativistic corrections.

\section{Non-relativistic Effective Field Theory for Atoms}
An atom is characterized by the following well-separated scales: $mv^2\ll mv\ll m \ll M_N$, with $v\sim \alpha\ll 1$
designating the typical velocity of electrons. All aspects of atomic physics can be adequately described by the
following effective lagrangian:
\begin{align}
	{\mathcal L}_{\rm atom}= {\cal L}_{\rm Maxwell}+ {\cal L}_{\rm NRQED} +{\cal L}_{\rm HNET},
	\label{Lag:atom:full}
\end{align}
where
\bseq
\label{Lag:4:items}
\begin{align}
	  & {\cal L}_{\rm Maxwell}= -\frac{1}{4} F_{\mu\nu}F^{\mu\nu}+\cdots,
	\\
	  & {\cal L}_{\rm NRQED}=\psi^\dagger \bigg\{iD_0+ {{\bf D}^2 \over 2m}+ \cdots\bigg\}\psi,
	\\
	  & {\cal L}_{\rm HNET}= N^\dagger i D_0 N+\cdots,
\end{align}
\eseq
where $\psi$ is the Pauli spinor field that annihilates an electron,
$N$ is the Dirac spinor field that annihilates a heavy nucleus at rest.
$D_\mu=\partial_\mu-ieA_\mu$ acts on electron field, $D_\mu=\partial_\mu+iZeA_\mu$ acts on nucleus field.
The nonrelativistic electron field is described by NRQED~\cite{Caswell:1985ui},
whereas the nucleus, whose role is solely
providing a static electric source, is treated in the {\it heavy nucleus effective theory} (HNET),
in a fashion analogous to the heavy quark effective theory (HQET)~\cite{Eichten:1989zv,Georgi:1990um}.
Here the nucleus is approximated by an
infinitely heavy, structureless point charge, and
we do not concern ourselves about the immaterial effect due to the magnetic moment of the nucleus.
Note \eqref{Lag:atom:full} only retains those operators at lowest order in
$v$ and $1/M_N$.

Eq.~\eqref{Lag:atom:full} is manifestly gauge invariant. However, a common practice to
tackle the nonrelativistic charged system is to impose the Coulomb gauge ${\bf\nabla}\cdot {\mathbf A}=0$,
where the instantaneous Coulomb photon can be cleanly separated from the dynamic transverse photons.
Since the $\bf A$ is suppressed with respect to $A^0$ in NRQED power counting,
it is a controlled approximation to drop all occurrences of $\bf A$, Eq.~\eqref{Lag:atom:full} then
essentially reduces into the following nonrelativistic Schr\"odinger field theory:
\begin{align}
	{\cal L}_{\rm Coul-Schr}=\psi^\dagger \Bigg\{iD_0+\frac{\bm{\nabla}^2}{2m}\Bigg\}\psi
	+N^\dagger i D_0 N+
	{1\over 2} \left(\nabla A^0\right)^2,
	\label{Lag:Coul:Schr:eff}
\end{align}
We solve the non-dynamic $A_0$ via the equation of motion $\partial_\mu\frac{\delta \mathcal{L}}{\delta\partial_\mu A_0}=\frac{\delta \mathcal{L}}{\delta A_0}$ and plug the solution into the Lagrangian. The resulting Lagrangian shows that the coupling of the theory is just an instantaneous Coulomb potential, which reads
\begin{eqnarray}
    {\cal L}_{\rm Coul-Schr}&=&\psi^\dagger \Bigg\{i\partial_0+\frac{{\bf\nabla}^2}{2m}\Bigg\}\psi
    +N^\dagger i \partial_0 N\nn\\
    &-&\int \textup{d}^3{\bf y}\  \left(\psi^\dagger\psi(x^0,{\bf x})-ZN^\dagger N(x^0,{\bf x})\right)\frac{\alpha}{|{\bf x-y}|}\nn\\
    &\times&\left(\psi^\dagger\psi(x^0,{\bf y})-ZN^\dagger N(x^0,{\bf y})\right),\nn\\
    \label{Lag:Coul:Schr:eff:second}
\end{eqnarray}
where $\alpha=\frac{e^2}{4\pi}$ is the fine-structure constant.
This theory may be dubbed Coulomb-Schr\"odinger effective lagrangian.
Obviously, the effect due to dynamical photon, such as Lamb shift~\cite{Pineda:1997ie},
will be inaccessible in \eqref{Lag:Coul:Schr:eff}.
Since the spin degree of freedom decouples in the nonrelativistic limit,
for simplicity we will replace the spinor fields $\psi$ and $N$ in
\eqref{Lag:Coul:Schr:eff} by complex scalar fields henceforth.

\section{Operator Product Expansion in Coulomb-Schr\"odinger EFT}
In order to deduce the coalescence behavior of the Coulomb-Schr\"odinger wave function of an atom,
we are motivated to examine how the product of the electron field $\psi$ and nucleus field $N$ scales
in short-distance, since in the field-theoretical context,
the wave functions can be viewed as the product of a string of spatially-nonlocal, yet equal-time $\psi$-fields and the
$N$-field, sandwiched between the vacuum and the bound state~\cite{Bodwin:1994jh}.

We start from the product of a single $\psi$ field and a $N$ field.
One may tentatively guess that, in the small $\mathbf{r}$ limit,
the operator product can be expanded as follows,
\begin{align}
	\psi(\mathbf{r}) N({\bf 0}) = [\psi N]({\bf 0})+\mathbf{r}\cdot [\bm{\nabla}\psi N]({\bf 0})+\cdots,
	\label{Naive:OPE:coordinate:space}
\end{align}
which is nothing but the Taylor expansion of $\psi(\mathbf{r})$ around the origin.
Note these two field operators are defined in equal time $t=0$.
The first operator in the right-hand side carries the $S$-wave quantum number,
the second is of $P$-wave type, and the ellipsis represent those irreducible-spherical-tensor operators
carrying two or more gradients.
We stress again that, it is crucial to introduce the HNET field to fulfill a valid OPE relation.

Incorporating the Coulomb interaction will modify the naive expectation of \eqref{Naive:OPE:coordinate:space}.
The highlight of this work is, as we will prove shortly,
that there exists an exact OPE relation in the EFT defined by \eqref{Lag:Coul:Schr:eff}: more terms in the parenthesis
\begin{align}
	\psi(\mathbf{r})N({\bf 0}) & =(1-mZ\alpha |\mathbf{r}|+\cdots)\,[\psi N]({\bf 0})
	\nn\\
	                           & + (1-mZ\alpha |\mathbf{r}|/2+\cdots)\mathbf{r}\cdot [{\bm\nabla}\psi N]({\bf 0})+\cdots,
	\label{Exact:OPE:coordinate:space}
\end{align}
where $[\ldots]$ is used to denote the renormalized composite operator.

Before moving on, it is worth emphasizing some significant difference between
renormalizable and nonrenormalizable (effective) theories on application of OPE.
In the former case, the distance $|\bf x|$ between two operators could literally tend to 0,
and the radiative corrections to each Wilson coefficient in the naive OPE series
bears the form of $\ln^n |\mathbf{r}|$.
By contrast, since \eqref{Lag:Coul:Schr:eff} only has a limited range of applicability,
the smallest distance of $|\mathbf{r}|$ one can probe is of order $1/m$, the inverse of the
UV cutoff of the Coulomb-Schr\"odinger EFT. The additional corrections to
each Wilson coefficient can be linear in $|\mathbf{r}|$, balanced by the Bohr radius
$a_0\equiv (mZ\alpha)^{-1}$.

The OPE in the momentum space turns out to be also useful:
\begin{align}
	  & \widetilde\psi({\bf q})N({\bf 0}) \equiv \int d^3 \mathbf{r}
	e^{-i {\bf q}\cdot \mathbf{r}} \psi(\mathbf{r})N({\bf 0})
	\nn \\
	  & = \frac{8\pi Z\alpha m}{{\bf q}^4} [\psi N]({\bf 0})-\frac{16 i \pi Z\alpha m}{{\bf q}^6}\bm{q}
	\cdot [{\bm\nabla}\psi N] ({\bf 0})+\cdots.
	\label{Exact:OPE:momentum:space}
\end{align}
This expansion is valid provided that $1/a_0 \ll |{\bf q}| \lesssim m$.
With the aid of the following rudimentary Fourier integrals,
\bseq
\begin{align}
	  & \int \!\! {d^3\mathbf{q}\over (2\pi)^3}\,\frac{e^{i\mathbf{q}\cdot\mathbf{r}}-1}{\mathbf{q}^4}
	=-{1\over 8\pi} |\mathbf{r}|,\label{Fourier1}
	\\
	  & \int\!\! {d^3\mathbf{q}\over (2\pi)^3}\,  {e^{i\mathbf{q}\cdot\mathbf{r}} - i \mathbf{q}\cdot \mathbf{r} \over \mathbf{q}^6}\,\mathbf{q}
	= -{i \over 32\pi}|\mathbf{r}| \mathbf{r},
\end{align}
\eseq
one readily recovers the coordinate-space OPE \eqref{Exact:OPE:coordinate:space}.
Note that one has to subtract the contribution
arising from the local composite operators,
to sweep IR divergences encountered in Fourier transform~\cite{Collins1984}.

\section{Proof of Operator Product Expansion in Coulomb-Schr\"odinger field theory}
The goal is to prove the momentum-space version of OPE, \eqref{Exact:OPE:momentum:space}, to all orders in $Z\alpha$.
The proof here is similar in spirit, but technically considerably simpler,
than that for the renormalizable quantum field theories~\cite{Zimmermann:1970,Collins1984}.
We note that Hofmann {\it et al.}~\cite{Hofmann:2013oia} recently proved an OPE relation governing the
electron-electron coalescence behavior starting from the Hamiltonian of the jellium model,
but essentially in a nonperturbative fashion.
Our proof is entirely based on perturbation theory.
Moreover, here the HNET field is introduced to
define an OPE relation.

Operator product expansion is an operator equation, which is insensitive to the presence of other fields or
external states. We take this freedom to investigate the following connected Green's functions:
\bseq
\begin{align}
	  & \Gamma\left({\bf{q}};\,{\bf{p}},E\equiv k^0+p^0\right) \equiv
	\int \! d^4y \, d^4z\  e^{-ip\cdot y-ik\cdot z} \langle 0 |T \{\widetilde\psi({\bf q})N({\bf 0})\psi^\dagger (y) N^\dagger (z)\}|0\rangle_{\rm amp},
	\label{Def:four-point:Green:function}
	\\
	  & \Gamma_S({\bf{p}},E) \equiv
	\int \!\! d^4y  d^4z\, e^{-ip\cdot y-ik\cdot z} \langle 0 \vert T\{[\psi N]({\bf 0})\psi^\dagger (y) N^\dagger (z)\} \vert 0\rangle_{\rm amp},
	\label{Def:Gamma:S}
	\\
	  & {\bm\Gamma}_P({\bf{p}},E) \equiv
	\int \!\! d^4y  d^4z\, e^{-ip\cdot y-ik\cdot z}\langle 0 |T\{[{\bf\nabla}\psi N]({\bf 0}) \psi^\dagger(y) N^\dagger (z)\}|0\rangle_{\rm amp},
\end{align}
\label{Def:three:amputated:Green:functions}
\eseq
where $p^\mu$, $k^\mu$ can be chosen as arbitrary 4-momenta that are much smaller than $m$. {In this nonrelativistic theory, the particle number must be conserved, so the above Green's functions are sufficient to draw a complete OPE relation. }
The subscript ``amp'' in \eqref{Def:three:amputated:Green:functions} implies that the
external propagators carrying soft momenta $p$ and $k$ get amputated.

The strategy is to verify that, order by order in $Z\alpha$, the large $|{\bf q}|$ behavior
of the Green's function $\Gamma$ does possess the following factorized structure\footnote{{Formally we take the inserted momentum $\vb{q}$ to infinity, but in the context of NREFT, such a quantity cannot exceed the electron mass $m$. What this limit actually means is to take $\vb{q}$ to a value where it is much larger than the inverse of Bohr radius $1/a_0$ but smaller than $m$ in the meantime. Similarly, in coordinate space, we take $r\to0$ formally, but in fact $r$ takes a value smaller than $1/m$ but much larger than $a_0$. }}:
\begin{align}
	  & \Gamma({\bf{q}};{\bf{p}},E) \xrightarrow{{\vb{q}\to\infty}}  \frac{8\pi mZ\alpha}{{\bf q}^4}
	\Gamma_S({\bf{p}},E)-\frac{i16\pi mZ\alpha}{{\bf q}^6}\bf{q} \cdot \bm{\Gamma}_P({\bf{p}},E) +\cdots.
	\label{OPE:Green:function:level}
\end{align}
\begin{figure}[tbh]
	\centering
	\includegraphics[width=0.2\textwidth]{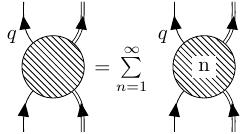} 
	\caption{
		$\Gamma=\sum_{n=1}^\infty \Gamma^{(n)}$.
		\label{Fig:Coulomb:ladder:sum}}
\end{figure}

Due to the specific causal structure of the propagators associated with
electron, nucleus, and Coulomb photon,
\begin{align}
	D_e(q^0,\bm{q})=
	{i\over q^0-{{\bf q}^2\over 2m}+i\epsilon},\;
	D_N(q^0)=\frac{i}{q^0+i\epsilon},\;
	D_C({\bm{q}})= {i\over {\bf q}^2},
	\nn
\end{align}
only those Coulomb ladder diagrams (crossed ladder, self-energy and vertex-correction type diagrams yield null results)
contribute to $\Gamma$, $\Gamma_S$ and $\bm{\Gamma}_P$, as the virtue of nonrelativistic Coulomb-Schr\"odinger field theory.
One can express $\Gamma=\sum_{n=1}^\infty \Gamma^{(n)}$, with $n$ signaling the number of exchanged Coulomb ladders
(the number of loops is then $n-1$). This topology is pictured in Fig.~\ref{Fig:Coulomb:ladder:sum},
with single and double solid lines representing the electron and nucleus, respectively.
$\Gamma_S$ and $\bm{\Gamma}_P$ also bear the same ladder structures,
thereby one can introduce $\Gamma_S^{(n)}$ and $\Gamma_P^{(n)}$ in a similar fashion.
It is worth mentioning that, to all orders in $Z\alpha$, the Green's functions
$\Gamma$, $\Gamma_S$ and $\Gamma_P$ are UV finite.

Explicitly, the $n$-ladder contribution to $\Gamma$ reads
\begin{align}
	\Gamma^{(n)} & = (Z e^2)^n \int\frac{d q^0}{2\pi} D_e(q) D_N(E-q^0)
	\left(\prod_{i=1}^{n-1} D_e(l_i) D_N(E-l_i^0)D_C({\mathbf{l}_i-\mathbf{l}_{i-1}})\right) D_C({\mathbf{l}_{n-1}-\mathbf{q}})
	\nn\\
	             & = (Z e^2)^n D_e(E;\mathbf{q})
	\label{Gamma:n:definition}
	\int \prod_{i=1}^{n-1}\frac{d^3 \mathbf{l}_i}{(2\pi)^3} \left(\prod_{i=1}^{n-1} D_e(E,\mathbf{l}_i)D_C({\mathbf{l}_i-\mathbf{l}_{i-1}})\right)
	 D_C({\mathbf{l}_{n-1}-\mathbf{q}}),
\end{align}
where $l_0\equiv p$. Note in the final expression, the upper-right HNET propagator in Fig.~\ref{Fig:Coulomb:ladder:sum}
disappears after integration over $q^0$,
as a consequence of Fourier-transforming the {\it equal-time} product of the $\psi$ and $N$ fields
in \eqref{Def:four-point:Green:function}. Interestingly, the hard momentum ${\bf q}$ injected
from the upper-left electron line exits through the upper-right HNET leg,
which is however of null impact since the HNET propagator is insensitive to
the residual nucleus three-momentum.
\begin{figure}[tbh]
	\centering
	\includegraphics[width=0.25\textwidth]{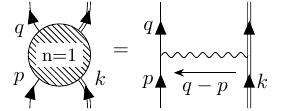}
	\caption{
		The tree-level amputated Green's function $\Gamma^{(1)}$.
		\label{Fig:One:ladder:tree}}
\end{figure}

Let us start with the lowest-order $\Gamma^{(1)}$, as depicted in Fig.~\ref{Fig:One:ladder:tree}. According to
\eqref{Gamma:n:definition}, we find
\begin{align}
	\Gamma^{(1)} & = Ze^2(-2m)\frac{i}{\bm{q}^2-2mE-i\epsilon}\frac{i }{|\bm{q}-\bm{p}|^2}
	\nonumber\\
	             & \xrightarrow{{\vb{q}\to\infty}} \frac{8\pi  mZ\alpha}{{\bf q}^4}+\frac{16\pi mZ\alpha}{{\bf q}^6}{\bf q}\cdot {\bf p}+\cdots,
	\label{Gamma:tree}
\end{align}
Since $|{\bf p}|,\,E \ll m$, we expand $\Gamma^{(1)}$ to sub-leading order in $1/|{\bf q}|$.
At lowest-order in $Z\alpha$, two Green's functions embedded with $S$ and $P$-wave local operators
can be readily evaluated, $\Gamma^{(0)}_S=1$ and
${\bm\Gamma}^{(0)}_P= i{\bf p}$. From \eqref{Gamma:tree} we thus immediately recognize that $\Gamma^{(1)}$ indeed possess
the factorized pattern as indicated in \eqref{OPE:Green:function:level}.
The extracted Wilson coefficients for $S$ and $P$-wave operators are compatible with
those given in \eqref{Exact:OPE:momentum:space}.

\begin{figure}[tb]
	\centering
	\includegraphics[width=0.4\textwidth]{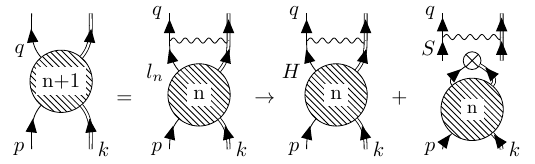}
	\caption{Reexpressing $\Gamma^{(n+1)}$ as the convolution of $\Gamma^{(n)}$ with
		one additional Coulomb photon together with the electron and nucleus propagator.
		The labels $H$ and $S$ indicate whether the loop momentum ${\bf l}_n$
		is hard ($\sim {\bf q}$) or soft  ($\ll m$).
		\label{Fig:Reexpress:N:ladder}}
\end{figure}

The challenge is to prove that the factorized form \eqref{OPE:Green:function:level}
persists to higher order in $Z\alpha$.
That is,  the Wilson coefficients are not subject to any further radiative corrections.
For this purpose, we resort to the method of induction. To keep everything as simple as possible, let us
first concentrate on the $S$-wave piece. Let us presume the asymptotic value of $\Gamma$ indeed obeys
the pattern specified in \eqref{OPE:Green:function:level} up to the order-$n$.
As indicated from Fig.~\ref{Fig:Reexpress:N:ladder}, we can reexpress
$\Gamma^{(n+1)}$ in terms of the one-loop integration of the $\Gamma^{(n)}$ together with one
additional Coulomb ladder together with electron and nucleus propagators,
where the momentum of the internal electron line is labeled by ${\bf l}_n$.
For the hard loop momentum, ${\bf l}_n\sim {\bf q}\lesssim m$, we can resort to the OPE assumption
$\Gamma^{(n)}({\bf l_n}; {\bf p}, E)\propto 1/{\bf l}_n^4$.
Power-counting the electron, Coulomb photon propagators and integration measure, one finds that
the net contribution from the hard-loop regime to $\Gamma^{(n+1)}$ scales as $|{\bf q}|^{-2-2+3-4}\sim |{\bf q}|^{-5}$,
and is thus power-suppressed and can be neglected.
For the soft-loop regime, ${\bf l}_n \sim {\bf p} \sim {\bf k} \ll m$,
the hard momentum ${\bf q}$ can only flow from the upper-left
external electron line into the upmost Coulomb photon, and exit through the external upper-right nucleus line.
It is then legitimate to expand the integrand in powers of $|{\bf p}|/|{\bf q}|$. Consequently,
as indicated in the rightmost diagram in Fig.~\ref{Fig:Reexpress:N:ladder}, one can dissect
$\Gamma^{(n+1)}$ into a high-energy factor multiplying a low-energy part, where the former exactly
corresponds to the tree-level Wilson coefficient extracted from \eqref{Gamma:tree},
and the latter can be identified with the Green's function containing the $S$-wave local operator, $\Gamma^{(n)}_S$ in \eqref{Def:Gamma:S},
which is nonanalytic function of ${\bf p}$ and $E$.

The similar analysis can be extended to include the $P$-wave operator in the right-hand side of \eqref{OPE:Green:function:level}.
One can further continue to include higher partial-wave operators that contain two or more gradients acting on $\psi$.
{By the above analysis, one can determine the leading components of the OPE relation. Here, the term "leading" has a dual meaning: it refers to the leading $q$ coefficient of the leading operator for each partial wave, specifically the operator with the minimal number of derivatives in each partial wave. To obtain the complete OPE relation, further but similar analyses are required, where the hard $q$ flows down to subsequent loops and so on, as shown in Fig.~\ref{Fig:Reexpress:N:ladder}. Additionally, for operators with higher orders of derivatives, the loop integral must be expanded to higher orders of $q^{-1}$ in each case of $q$ flow to obtain the Wilson coefficients. However, the focus of this paper is to understand the universal behavior of hydrogen wavefunctions at the first order of their Taylor expansion. Thus, we are content with the leading-order analysis presented here. }

%

\section{Application of OPE to Hydrogen-like atoms}
The OPE relations \eqref{Exact:OPE:coordinate:space} and \eqref{Exact:OPE:momentum:space}
provide a lucid way to understand the {\it universal} electron-nucleus coalescence
behavior of wave function for any {\it arbitrary} atom.
Nevertheless, for the sake of illustration, we choose the simplest and extremely well-known case, {\it e.g.},
the wave functions of hydrogen-like atoms to test our formalism.
The corresponding wave functions can be identified with the following spatially nonlocal matrix element:
\begin{align}
	\Psi_{nlm}(\mathbf{r})=\left\langle 0\left\vert\psi(\mathbf{r})N({\bf 0})\right\vert nlm \right\rangle,
	\label{Hydrogen:wave:function}
\end{align}
where $n$, $l$, $m$ refer to radial, orbital, and magnetic quantum numbers, respectively.

Sandwiching \eqref{Exact:OPE:coordinate:space} between the vacuum and the $S$-wave or $P$-wave states,
picking up the operator carrying the corresponding orbital quantum number,
after some algebra we predict that the radial wave functions near the origin become approximately
\bseq
\begin{align}
	  & R_{n0}(r) \xrightarrow{{r\to0}} R_{n0}(0)\left(1- {r\over a_0}\right),
	\\
	  & R_{n1}(r) \xrightarrow{{r\to0}}  r R^\prime_{n0}(0) \left(1- {r\over 2 a_0} \right),
\end{align}
\label{Hydrogen:radial:small:x}
\eseq
where the shorthand $r=|\mathbf{r}|$ is used.
Note \eqref{Hydrogen:radial:small:x} is true for any $n$, regardless of being
discrete or continuum label. One can explicitly confirm these relations by expanding the exact hydrogen
atom wave functions near the origin.
Reassuringly, the universal behavior is found to be {\it violated} at order-$r^2$ for each partial wave,
which confirms our anticipation that OPE must break down at this relative order.

A general relation concerning the wave functions near the origin for the hydrogen-like atoms,
valid for any $l$, was given by L\"owdin in 1954~\cite{Loewdin1954}:
\begin{align}
	R_{nl}(r)= \frac{r^l}{l!}\frac{d^l R_{nl}}{dr^l}(0)
	\left[1-{1\over l+1}{r\over a_0}+{\cal O}\left(r^2/a_0^2\right)\right],
	\label{Lowdin:formula:hydrogen}
\end{align}
which can be recast into an equivalent form,
$$
	\frac{d^lR_{nl}(r)}{dr^l} =\frac{d^lR_{nl}(0)}{dr^l}
	\left[1-{r\over a_0}+{\cal O}\left( r^2/a_0^2\right) \right].
$$
Obviously, our OPE predictions for $S$- and $P$-wave hydrogen-like atoms,
\eqref{Hydrogen:radial:small:x}, coincide with L\"owdin's relation
\eqref{Lowdin:formula:hydrogen}.

Eq.~\eqref{Exact:OPE:momentum:space} can also be applied to predict the universal behavior of
momentum-space wave functions in the large momentum limit:
\begin{align}
	\widetilde R_{nl}(q)= 2^{l+2}\frac{d^lR_{nl}(0)}{d x^l}\frac{(2\pi)^\frac{5}{2}}{a_0 q^{l+4}}
	\left[1+{\cal O}(1/(q a_0)^2)\right],
	\label{wave:function:large:q:scaling}
\end{align}
with $q\equiv |{\bf q}|$, and $n$ can again be either discrete or continuum label.
This universal behavior of the bound wave functions in momentum space
appears to be first noted in the famous text by Bethe and Salpeter~\cite{Bethe1977},
yet without any physical explanations. One can verify our OPE predictions by directly
expanding the exact momentum-space radial wave functions for both bound~\cite{Podolsky1929}
and continuum~\cite{Dolinskii1966} states for arbitrary $l$.

\section{Extension to two-electron coalescence}
Although electron-electron coalescence has been discussed by Hofmann with the background field technique and Bethe-Salpeter equation, we apply our perturbative method to such a problem to display the validity and efficiency of our method.

While this procedure is quite similar to electron-nucleus coalescence, the spin states of two different electrons must be explicitly labeled. 
The OPE in momentum space reads
\begin{align}
	  & n(\vb{q}) \equiv \int\dd^3\vb{r}e^{-i\vb{q}\cdot\vb{r}}\psi_a\pqty{\frac{\vb{r}}{2}}\psi_b\pqty{-\frac{\vb{r}}{2}}
	\nn \\
	  & = -\frac{4\pi m \alpha}{{\bf q}^4} [\psi_a \psi_b]({\bf 0})+\frac{8 i \pi m \alpha}{{\bf q}^6}\bm{q}
	\cdot [{\bm\nabla}\psi_a \psi_b] ({\bf 0})+\cdots.
	\label{Exact:OPE:momentum:space2}
\end{align}
where the P-wave operator is defined as
\begin{align}
	[{\bm\nabla}\psi_a \psi_b]\equiv \frac{[\psi_a\bm{\nabla}\psi_b]+[\psi_b\bm{\nabla}\psi_a]}{2}
\end{align}
The coordinate space counterpart is
\begin{align}
	\psi_a\pqty{\frac{\vb{r}}{2}}\psi_b\pqty{-\frac{\vb{r}}{2}}= & (1+m \alpha |\mathbf{r}|/2)\,[\psi_a \psi_b]({\bf 0})
	\nn\\
	                                                             & + (1+m \alpha |\mathbf{r}|/4)\mathbf{r}\cdot [{\bm\nabla}\psi_a \psi_b]({\bf 0})+\cdots,
	\label{Exact:OPE:coordinate:space2}
\end{align}
For proof of the above relations, see Appendix~\ref{App:B}.

Following we apply the OPE result \eqref{Exact:OPE:coordinate:space2} to the actual physical scenario. The atom wave functions $\Phi(\vb{R},\vb{r}_1,\dots,\vb{r}_n)$ is one example, in the case that the atomic number is larger than 1. When two electrons inside an atom approach each other, the S-wave wave function takes an asymptotic form\cite{Kato1957}
\begin{align}
	   \Phi(\vb{R},\vb{r}_1,\dots,\vb{r}_n)\notag &=\left\langle 0\left\vert\psi\pqty{\vb{r}_1}\dots\psi_a\pqty{ \vb{r}_a} \psi_b\left( \vb{r}_b \right)\dots\psi\pqty{\vb{r}_n}N\pqty{\vb{R}}\right\vert\text{atom}\right\rangle\\&\xrightarrow{\vb{r}_{ab}\to 0}\Phi(\vb{r}_{ab}=0)\pqty{1+m\alpha\abs{\vb{r}_{ab}}/2}
\end{align}
where $\vb{r}_{ab}$ is the relative vector between electrons a and b.

Another example is the jellium model mentioned in \cite{Hofmann:2013oia}, where electrons are bonded by a background field. The N-particle wave function $\Psi\pqty{-\frac{\vb{r}}{2},\uparrow;\frac{\vb{r}}{2},\downarrow;\vb{r}_3,\sigma_3;\dots}$, with given operator level definition, exhibits the same {factorized behavior and Wilson coefficient. }{We would like to add a few remarks here: Hofmann et al. conducted a very similar exploration in \cite{Hofmann:2013oia}, where the concept of OPE was employed to extract the short-distance behavior between electrons. Unlike our perturbative approach, they solved the Green's functions nonperturbatively using the Bethe-Salpeter equation for the jellium model and reached the same conclusion as ours in this section. However, in the present work, our primary focus is on deriving the electron-nucleus OPE from first principles with the implementation of HNET. As a result, we obtained OPE relations for all partial waves and explicitly demonstrated that the Wilson coefficients for electron-electron and electron-nucleus coalescence differ only by the reduced mass, a point briefly mentioned by \cite{Hofmann:2013oia} in a footnote.}

\section{Summary and Outlook}
The atomic physics and quantum chemistry are quite mature fields,
the central theme of which is to effectively solve the
Schr\"odinger equations for atoms and molecules. Knowing the true coalescence
behavior of wave functions provide important guidance for constructing the optimal trial wave functions.
While the two-particle and three-particle coalescence behavior are
known, it seems to be a formidable task to extend the traditional differential-equation-based approach to
infer the four or more particle coalescence behaviors.

This work approaches this old problem from a different and field-theoretical perspective.
Concretely, we have rigorously proved an operator product expansion relation
within the Coulomb-Schr\"odinger effective field theory. This OPE
relation can naturally explain Kato's cusp condition when an electron approaches the nucleus. {With its operator-based nature, this method can be implanted for any relevant physical quantity embedding two nonrelativistic fields when the short distance behavior is concerned. }
Moreover, our approach, which is based upon the systematic Feynman-diagrammatic technique,
can be straightforwardly generalized to deduce the
universal multi-particle coalescence behavior of atomic or molecular wave functions, by
constructing a multi-field OPE relation.
This information will provide invaluable guidance for constructing optimized trial wave functions
for large-$Z$ atoms. We hope to explicitly unravel the coalescence behavior for three
and four electrons coinciding with the nucleus in the future publication. 

Another interesting direction of advance is to incorporate relativistic corrections
into our Coulomb-Schr\"odinger EFT. Curiously, our OPE formalism appears to be able to
provide some crucial insight into settling a 90-year-old puzzle related to
the relativistic Klein-Gordon and Dirac wave equation with a Coulomb potential,
namely, what is the mechanism responsible for the {\it universal}
weak divergence of wave function near the origin for the $S$-wave hydrogen atom is~\cite{Huang:2018ils,Huang:2019hdj}.

\appendix
\section{Wavefunctions of Hydrogen-like Atoms}
The Hamiltonian for Hydrogen is given by
\begin{align}
	H=-\frac{\bm{\nabla}^2}{2m}-\frac{Z\alpha}{r}
\end{align}
The Schr\"odinger equation with Coulomb potential is
\begin{align}
	\pqty{-\frac{\bm{\nabla}^2}{2m}-\frac{Z\alpha}{r}}\Psi(\vb{r})=E\Psi(\vb{r})
\end{align}
Separating the radial and angular part we have
\begin{align}
	  & \Psi_{xlm} (r,\theta,\phi)=R_{xl} (r) Y_{lm}(\theta,\phi); \\&\tilde\Psi_{xlm} (q,\theta_q,\phi_q)=\tilde R_{xl} (q) Y_{lm}(\theta_q,\phi_q);
\end{align}
where $x$ indicates the discrete or continuum index. The coordinate space solution to Schr\"odinger hydrogen is given by\cite{Bethe1977}
\begin{widetext}
	\begin{align}
		  & R_{nl} (r)       =\frac{2a_0^{-\frac{3}{2}}}{n^2(2l+1)!}\sqrt{\frac{(n+l)!}{(n-l-1)!}}e^{-\frac{ r}{a_0n}}\pqty{\frac{2 r}{a_0n}}^l{}_1F_1(l+1-n,2l+2,\frac{2  r}{a_0n}),\ \ \ E_n=-\frac{1}{2n^2 m a_0^2}                             \\
		  & R_{k l} (r) =\sqrt{\frac{2}{\pi}}\frac{k}{(2l+1)!} e^{\frac{\pi}{2a_0k}}|\Gamma(l+1-\frac{i}{a_0k})|e^{-ik r}\pqty{2k r}^l{}_1F_1(l+1+\frac{i}{a_0k},2l+2, 2ikr),\ \ \ E_{k}=\frac{k^2}{2m},
		\intertext{{after expanding the above wave functions, we obtain exactly the short distance behavior in eqn. \eqref{Exact:OPE:coordinate:space} and \eqref{Lowdin:formula:hydrogen}. We also have the }momentum space solution\cite{Podolsky1929,Dolinskii1966}}
		  & \tilde R_{nl} (q)       =\pi^\frac{5}{2}2^{2l+\frac{11}{2}}\pqty{ a_0n}^{3/2}l!\sqrt{\frac{n(n-l-1)!}{(n+l)!}}\frac{\xi^l}{(\xi^2+1)^{l+2}}C^{l+1}_{n-l-1}\pqty{\frac{\xi^2-1}{\xi^2+1}}                                               \\
		  & \tilde R_{k l} (q)       =-\frac{16\pi^2\eta e^{-\pi\eta/2}k^2(qk)^l}{(q^2+k^2)^{1+l+i\eta}}\bqty{\frac{\Gamma{(1+l+i\eta)}}{(1/2)_{l+1}}} {}_2F_1(\frac{2+l+i\eta}{2},\frac{1+l+i\eta}{2};l+3/2;\frac{4k^2q^2}{(q^2+k^2)^2})\nonumber \\&\hspace*{350pt}\times \lim_{\gamma\rightarrow 0}\bqty{q^2-(k+i\gamma)^2}^{-1+i\eta}
	\end{align}
\end{widetext}
where $C_l^m(x)$ is the Gegenbauer C function, $(a)_n$ stands for the Pochhammer symbol, $a_0=(mZ\alpha)^{-1}$ is the Bohr radius of the atom, $\xi\equiv qa_0n$, $\eta=-\frac{1}{ a_0k}$. {After expanding over large-$q$, we obtain the short distance behavior in eqn. \eqref{Exact:OPE:momentum:space} and \eqref{wave:function:large:q:scaling}. }
\section{OPE for Electron-electron Coalescence\label{App:B}}
To prove the OPE relations in \eqref{Exact:OPE:momentum:space2} and \eqref{Exact:OPE:coordinate:space2}, we follow the same methodology with electron-nucleus coalescence,starting with defining several four-point connected amputated Green's functions 
\begin{align}
	  & \Gamma_e\left({\bf{q}};\,{\bf{p}},{\bf{k}},p^0\right) \equiv
	\int \! d^4y \, d^4z\  e^{-i(p+k)\cdot y-i(p-k)\cdot z}
	\langle 0 |T \{n(\vb{q})\psi_a^\dagger (y) \psi_b^\dagger (z)\}|0\rangle_{\rm amp},
	\label{Def:four-point:Green:function2}
	\\
	  & \Gamma_{eS}({\bf{p}},{\bf{k}},p^0) \equiv
	\int \!\! d^4y  d^4z\, e^{-i(p+k)\cdot y-i(p-k)\cdot z}
	\langle 0 \vert T\{[\psi_a \psi_b]({\bf 0})\psi_a^\dagger (y) \psi_b^\dagger (z)\} \vert 0\rangle_{\rm amp},
	\label{Def:Gamma:S2}
	\\
	  & {\bm\Gamma}_{eP}({\bf{p}},{\bf{k}},p^0) \equiv
	\int \!\! d^4y  d^4z\, e^{-i(p+k)\cdot y-i(p-k)\cdot z}
	\langle 0 |T\{[{\bf\nabla}\psi_a \psi_b]({\bf 0}) \psi_a^\dagger(y) \psi_b^\dagger (z)\}|0\rangle_{\rm amp}.
\end{align}
The leading order Wilson coefficient comes from the tree diagram of $\Gamma$ in momentum space:
\begin{align}
	  & \phantom{=}\includegraphics{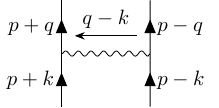}\label{eeTree}                                                                         \\&=
	-e^2\int\frac{\dd q^0}{2\pi}\frac{i}{p^0+q^0-\frac{\abs{\vb{q+p}}^2}{2m}+i\epsilon}\frac{i}{p^0-q^0-\frac{\abs{\vb{p-q}}^2}{2m}+i\epsilon}\frac{i}{\abs{\vb{q-k}}^2}\notag\\
	  & =e^2\frac{-2m}{\abs{\vb{q+p}}^2+\abs{\vb{q-p}}^2-4mp^0}\frac{1}{\abs{\vb{q-k}}^2}                \notag \\
	  & \xrightarrow{\vb{q}\to \infty}-\frac{4\pi m\alpha}{\vb{q}^4}-\frac{8\pi m \alpha \vb{q}\cdot\vb{k}}{\vb{q}^6}+\cdots\notag
\end{align}
Employing the same Fourier transform technique in \eqref{Fourier1}, we arrive at the Wilson coefficients in \eqref{Exact:OPE:coordinate:space2}. {It's worth mentioning that the above diagram shares similar pole structure in the $q^0$-plane with the tree diagram in Fig~\ref{Fig:One:ladder:tree} and \eqref{Gamma:tree}. However, \eqref{eeTree} replaced the HNET propagator in \eqref{Gamma:tree} with a NRQED propagator, where extra three-momentum $\vb{p}-\vb{q}$ was introduced. After the contour integral over $q^0$, the extra NRQED propagator brings a second $\vb{q}^2$ term in the denominator of the first fraction. This is why the leading order Wilson coefficient is half the size of the one in the electron-nucleus case. This conclusion concurs with the claim in \cite{Hofmann:2013oia}, which states that their electron-electron result, counting the reduced mass, can be adapted for a hydrogen atom. From a Schr\"odinger equation point of view, this is obvious: these two systems only differ by the reduced mass. But from an EFT point of view, the HNET field and the NRQED field are intrinsically different. The nucleus hardly recoils when struck by incoming momenta, in contrast to the electron. Nonetheless, at leading order, the above contour integral ensures the Wilson coefficients of electron-electron and electron-nucleus coalescence are indeed identical except for the reduced masses. }

This result agrees with Kato's result \eqref{Kato:cusp:condition} and others' as well. We can see that applying this perturbative method is much simpler than solving a Bethe-Salpeter equation, without losing any physical content.
\begin{figure}[H]
	\centering
	\includegraphics[width=0.4\textwidth]{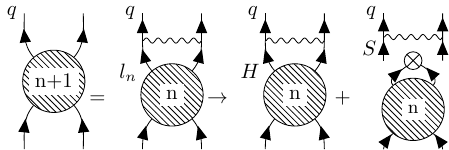}
	\caption{Reexpressing $\Gamma_e^{(n+1)}$ as the convolution of $\Gamma_e^{(n)}$ with
		one additional Coulomb photon together with the electron and nucleus propagator.
		The label $H$ and $S$ indicate whether the loop momentum ${\bf l}_n$
		is hard ($\sim {\bf q}$) or soft  ($\ll m$).}
	\label{Fig:Reexpress:N:ladder2}
\end{figure}

An all-order proof can be provided via the factorized structure in Fig~\ref{Fig:Reexpress:N:ladder2}. To achieve this, we must first integrate out all temporal components of $\Gamma_e$. With $l_0\equiv k$ we have
\begin{align}
	\Gamma_e^{(n)} & = (- e^2)^n \int\frac{d q^0}{2\pi} D_e(p+q) D_e(p-q)
	              \int \prod_{i=1}^{n-1}\frac{d^4 l_i}{(2\pi)^4}
	\left(\prod_{i=1}^{n-1} D_e(p+l_i) D_e(p-l_i)D_C({\mathbf{l}_i-\mathbf{l}_{i-1}})\right)
	             D_C({\mathbf{l}_{n-1}-\mathbf{q}})
	\nn\\
	               & = (- e^2)^n \frac{i}{2p^0-\frac{\vb{p}^2+\vb{q}^2}{m}+i\epsilon}
	\label{Gamma:n:definition2} \int \prod_{i=1}^{n-1}\frac{d^3 \mathbf{l}_i}{(2\pi)^3} \left(\prod_{i=1}^{n-1} \frac{i}{2p^0-\frac{\vb{p}^2+\vb{l}_i^2}{m}+i\epsilon}D_C({\mathbf{l}_i-\mathbf{l}_{i-1}})\right)
	 D_C({\mathbf{l}_{n-1}-\mathbf{q}}),
	\nn
\end{align}
which is quite similar to \eqref{Gamma:n:definition} in the large-$\vb{q}$ limit except for an $1/2$ factor. We can use the same hard-soft formalism to justify such factorization.

\begin{acknowledgments}
	{\noindent\it Acknowledgment.}
	We are grateful to Jia-Yue Zhang for bringing some references to our attention.
	This work is supported in part by the National Natural Science Foundation of China under Grants No.~11475188 and No. 11925506.
\end{acknowledgments}

\end{document}